\documentstyle[12pt,graphicx]{article}
\title{
Analogue surface gravity near the QCD chiral phase transition
 }
\author{
Neven Bili\'c and Dijana Toli\'c\\
Rudjer Bo\v{s}kovi\'{c} Institute, \\
P.O.\ Box 180, 10001 Zagreb, Croatia \\
E-mail: bilic@thphys.irb.hr, dijana.tolic@irb.hr
}
\date{\today}
\begin{document}
\maketitle
\begin{abstract}

 Using the formalism of relativistic acoustic geometry we study the 
expanding chiral fluid in the regime of broken chiral symmetry near the QCD chiral phase transition
temperature $T_{\rm c}$. The dynamics of pions below $T_{\rm c}$
is described by the equation of motion for a massless scalar field propagating  in 
curved spacetime similar to an open FRW universe.
 The metric tensor
depends locally on the
soft pion dispersion relation
and the four-velocity
 of the fluid.
In the neighbourhood of the critical point 
an analogue trapped region forms with
the analogue trapped horizon as its boundary.
We show that the associated 
  surface gravity
 diverges near the critical point as 
$\kappa \sim (T_{\rm c}-T)^{-1}$. Hence,
if the horizon forms close to the critical temperature
the analogue Hawking temperature may be comparable with or even larger than
the background fluid temperature.
\vspace{0.1in}
\end{abstract}

\section{Introduction}

\label{introduction}
Analog gravity  models of general relativity seem promising routes
to providing laboratory tests of the foundation of
quantum field theory in curved spacetime
\cite{novello} (for a review and an extensive list of references, see \cite{barcelo}).
The models are based on kinematics of waves
propagating in
inhomogeneously flowing media.
Various aspects of these phenomena have been studied in acoustics
\cite{visser}
optics \cite{philbin}
and superfluidity
\cite{jacobson}.
In this paper we study the analogue gravity model based on massless pions
propagating in a hadronic fluid. 

Strongly interacting matter is described at the fundamental level by
a nonabelian gauge theory called
 quantum chromodynamics (QCD).  At large distances 
or small momenta, the QCD exhibits 
the phenomena of quark confinement
and chiral symmetry breaking.
Chiral symmetry breaking expresses 
the fact that massless quarks confined in hadrons appear
effectively as massive constituents with a dynamically generated mass of several 
hundred MeV.  At low energies, the QCD vacuum is characterized by 
a nonvanishing expectation value
\cite{shifman,harris}:
  $\langle \bar\psi\psi\rangle \approx$ (235 
MeV)$^3$,
the so-called quark condensate,
which describes the density of quark-antiquark pairs
found in the QCD vacuum and its nonvanishing value is the manifestation of chiral symmetry breaking.
The chiral symmetry is restored at finite temperature above
certain critical temperature $T_{\rm c}$ of the order of 150 MeV.
In the chirally broken phase, below $T_{\rm c}$  
the pions, although being
massless, propagate slower than light
\cite{pisarski2,son1,son2} with velocity approaching zero at the
critical temperature.
The pions propagate through the expanding fluid  
the velocity of which may well be comparable
with the speed of light.
Hence, it is very likely that the flow velocity will exceed the pion velocity
if the temperature of the fluid is below and close to the critical temperature.
A similar condition is also realized in cosmological models of expanding universe
in which the recession velocity of the expanding spacetime
beyond the Hubble horizon exceeds the speed of light.

In order to explore this analogy with cosmology more closely
we need to specify the velocity field of the flow in a given spacetime geometry
and the fluid parameters such as the density, temperature
and the velocity of pions propagating in the fluid.
For that purpose we exploit a simple boost invariant Bjorken type spherical expansion \cite{bjorken}
of the chiral fluid. A similar model has been studied some time ago in the context of 
disoriented chiral condensate \cite{lampert}.
Furthermore, the velocity of pions in the chiral fluid can be derived using a linear sigma model as 
an effective low energy model of strong interactions.
Then, the propagation of massless pions  provides a setting for a geometric  analogue of expanding spacetime.
Since the velocity of the fluid may exceed the velocity of pions it is conceivable that an apparent horizon
may exist with the associated analogue Hawking radiation.
 In this paper we study the analogue Hawking effect 
using the Kodama-Hayward definition of surface gravity \cite{hayward2}.

In the next two sections we describe our basic concepts  and in sec.~\ref{surface}
we derive an expression for surface gravity and study its  behaviour near the critical point.

\section{Dynamics of the chiral fluid}
\label{chiral}
In the following we assume that the chiral fluid undergoes a 
 boost invariant Bjorken-type spherical expansion.
A Bjorken-type expansion is a simple and very useful hydrodynamic model that 
reflects the boost invariance of the deep inelastic scattering  in high energy collisions.
The original model [11] was introduced to describe the longitudinal expansion only.
A more realistic model of heavy ion collisions involves a transverse expansion superimposed on the longitudinal 
boost invariant expansion
\cite{dumitru-kolb}.
In order to
draw the analogy with cosmology,
here we consider a spherically symmetric 
 Bjorken expansion\footnote{From the high energy physics perspective, a spherical
expansion  is more appropriate for $e^+e^-$ collisions 
\cite{cooper}
because in this case the jets are produced with
no directional preference.}
 which is invariant under
radial boosts. 
In this model the radial three-velocity in
 radial coordinates
$x^\mu=(t,r,\vartheta,\varphi)$ is  a simple function 
$v=r/t$. Then the four-velocity  
 is  given by
\begin{equation}
u^\mu= (t/\tau, r/\tau,0,0) ,
\label{eq144}
\end{equation}
where $\tau=\sqrt{t^2-r^2}$ is the 
proper time. With the substitution
\begin{eqnarray}
& &t=\tau \cosh y ,
\nonumber \\
& & r=\tau \sinh y ,
\label{eq147}
\end{eqnarray}
the radial velocity is expressed as
\begin{equation}
v=\tanh y ,
\label{eq246}
\end{equation}
and the four-velocity as
\begin{equation}
u^\mu=(\cosh y,\sinh y,0, 0).
\label{eq146}
\end{equation}
The substitution (\ref{eq147}) may be regarded as a coordinate transformation
from ordinary radial coordinates 
to  new coordinates
$(\tau,y,\vartheta,\varphi)$
in which
the flat background metric
takes the form
\begin{equation}
g_{\mu\nu}= {\rm diag} \left(1,  -\tau^2, -  
\tau^2\sinh^2\! y, - \tau^2\sinh^2\! y \sin^2\!\theta \right),
\label{eq108}
\end{equation}
and the velocity components become  $u^\mu=(1,0,0,0)$.
Hence,
the new coordinate frame is comoving.
The metric  corresponds to an FRW expanding cosmological model 
with cosmological scale  $a=\tau$ and negative spatial curvature.
 The transformation (\ref{eq147}) maps the spatially flat
Minkowski spacetime into an expanding FRW spacetime with cosmological scale $a=\tau$ and negative spatial 
curvature. The resulting flat spacetime with metric (\ref{eq108}) is  known in cosmology 
as the {\it Milne universe} \cite{milne}.

The temperature of the expanding chiral fluid, to a good approximation, is proportional to $\tau^{-1}$.
This follows from the fact that the chiral matter is dominated by massless pions,
and hence, the density of the fluid 
may be approximated by the density $\rho=(g\pi^2/30) T^4$ of an ideal massless boson gas 
\cite{landau}.
 Using this and the energy-momentum conservation one finds
\begin{equation}
T=\frac{c_0}{\tau},
\label{eq007}
\end{equation}
where the constant $c_0$  may, in principle, be fixed from 
the phenomenology of high energy collisions.

Next we derive the pion velocity in the chiral fluid.
The dynamics of mesons in  a medium  is described by
a chirally symmetric Lagrangian of the form
\cite{son1,son2,bilic2}
\begin{equation}
{\cal{L}} =
 \frac{1}{2}(a\, g^{\mu\nu}
 +b\, u^{\mu}u^{\nu})\partial_{\mu} \varphi
 \partial_{\nu} \varphi
 - \frac{m_0^2}{2}
 \varphi^2
- {\lambda\over 4}
 (\varphi^2)^2 ,
 \label{eq1}
\end{equation}
where $u_{\mu}$ is the velocity of the fluid,
and $g_{\mu\nu}$ is the background metric.
 The mesons $\varphi\equiv (\sigma ,$
{\boldmath$\pi)$}  constitute
the $(\frac{1}{2},\frac{1}{2})$
 representation of the chiral SU(2)$\times$SU(2).
The parameters
 $a$ and $b$ depend  on the local temperature $T$
 and on the parameters of the model $m_0$ and $\lambda$
 and may be calculated
in perturbation theory.
At zero temperature the medium is  absent in which case $a=1$ and $b=0$.

 If $m_0^{2} < 0$ the chiral
symmetry will be spontaneously broken.
At the classical level, the $\sigma$  field develops a
nonvanishing expectation value such that
$ \langle \sigma \rangle =
f_{\pi}$.
At nonzero temperature the expectation value  
$\langle \sigma \rangle$, usually referred to as the chiral condensate,  is  temperature dependent
and vanishes at the chiral transition point.
Hence, the quantity  $\langle \sigma\rangle$ serves as an order parameter. 
 For temperatures below the
chiral transition point the meson masses are given by
\begin{equation}
m_{\pi}^2 =  0\, ; \;\;\;\;\;\;
m_{\sigma}^2 = 2\lambda \langle\sigma\rangle^{2}  ,
\label{eq43}
\end{equation}
in agreement with the Goldstone theorem.
The temperature dependence of the chiral condensate $\langle\sigma\rangle$
is obtained by
minimizing the thermodynamical potential
$\Omega=-(T/V) \ln Z$
with respect to
 $\langle\sigma\rangle$
at fixed temperature $T$.
At one loop order the solution to  the extremum condition
 as a function of temperature exhibits a weak first-order
 phase transition \cite{bilic1,rod}.
 However,  Pisarski and Wilczek have shown on general grounds
  that  the phase transition in SU(2)$\times$SU(2) chiral models should be of second order \cite{pis}.
  Hence, it is generally believed that a  first-order
 phase  transition  in this case is an artifact of
   the one loop approximation.
      Two loop calculations \cite{baacke} make an improvement and 
   confirm the general analysis of  \cite{pis}.
 
Propagation of pions is governed by the equation of motion  
\begin{equation}
\frac{1}{\sqrt{-g}}
\partial_{\mu}
\left[
{\sqrt{-g}}\,
( a\, g^{\mu\nu}+ b\,
u^{\mu}
u^{\nu}) \partial_{\nu}\mbox{\boldmath{$\pi$}}\right]
+V(\sigma,
\mbox{\boldmath{$\pi$}})
\mbox{\boldmath{$\pi$}}=0,
\label{eq013}
\end{equation}
which follows from 
the effective  Lagrangian obtained from (\ref{eq1}) by  spontaneous chiral symmetry breaking.
The quantity  $V$  in (\ref{eq013}) is the interaction potential
the form of which is irrelevant for our
consideration.
From (\ref{eq013}) we obtain the pion velocity squared as
\cite{bilic2}
\begin{equation}
c_{\pi}^2=\frac{a}{a+b} .
\label{eq015}
\end{equation}
The parameters $a$ and $b$  at nonzero temperature may be derived from the
finite temperature self energy  $\Sigma(q,T)$ of the pion
in the limit when the external momentum
$q$ approaches 0.
For a flat background geometry $g_{\mu\nu}=\eta_{\mu\nu}$,
the inverse pion propagator $\Delta^{-1}$  is given by
\begin{equation}
 \Delta^{-1}=a q^{\mu}q_{\mu}
 +b(q^{\mu}u_{\mu})^2 -m_{\pi}^2,
  \label{eq200}
\end{equation}
or in comoving frame, i.e., in a reference frame in which
$u^{\mu}=(1,0,0,0)$, 
\begin{equation}
 \Delta^{-1}=(a+b)q_0^2
 -a \mbox{\boldmath $q$}^2.
  \label{eq202}
\end{equation}
Hence, the parameters $a$ and $b$,
  can be expressed in terms of second derivatives of
 $\Sigma(q,T)$ with respect to $q_0$ and $q_i$
evaluated at $q^{\mu}=0$.

The pion velocity  at nonzero  temperature
was  calculated at one loop level by Pisarski and Tytgat
in the low temperature approximation
\cite{pisarski2}.
At one loop level the only diagram that
gives a nontrivial $q$-dependence of $\Sigma(q,T)$ is the bubble
diagram and the
calculation may be performed for the hole range of temperatures below the 
chiral critical point \cite{son2,bilic2}.
As we are particularly interested in the behaviour near the critical point 
of the chiral phase transition
we make use of the exact results 
near the critical temperature based on scaling and universality analysis of Son and Stephanov \cite{son1}.
In their notation our parameters $a$ and $b$ are expressed as
\begin{equation}
a =\frac{f_s^2}{\langle \sigma\rangle^2}
 ; \hspace{1cm} a+b =\frac{f_t^2}{\langle \sigma\rangle^2} ,
\label{eq351}
\end{equation}
where $f_s$ is equal to the temperature dependent pion decay constant and
$f_t^2$ coincides with the isospin susceptibility $\chi_{I5}$.
The chiral condensate $\langle \sigma\rangle$ up to a factor of the order $f_\pi^2$ equals
the quark condensate $\langle \bar\psi\psi\rangle$ at zero quark masses.
In the limit $T\rightarrow T_{\rm c}$ the quantity $f_t^2$ goes to a nonzero constant 
whereas, contrary to naive expectations, the scaling of $f_s$ turns out to be
  different than that of the order parameter $\langle \sigma\rangle$:
\begin{equation}
f_s^2  \sim (T_{\rm c}-T)^{(d-2)\nu}
 ; \hspace{1cm} \langle \sigma\rangle \sim (T_{\rm c}-T)^\beta ,
\label{eq251}
\end{equation}
where  $d$ is the number of  space dimensions and $\nu$ and $\beta$ are  positive 
critical exponents. 
 Hence, we obtain
\begin{equation}
a \sim (T_{\rm c}-T)^{(d-2)\nu -2\beta}
 ; \hspace{1cm} a+b \sim (T_{\rm c}-T)^{-2\beta} ,
\label{eq252}
\end{equation}
so the pion velocity scales as
\begin{equation}
c_\pi^2 \sim (T_{\rm c}-T)^{(d-2)\nu}, 
\label{eq253}
\end{equation}
as one approaches the critical temperature.
In $d=3$ dimensions, the critical exponents for  the O(4) universality class are $\nu=0.73$ and $\beta=0.38$  
 \cite{baker}.

\section{Chiral geometry}
\label{apparent}

We now introduce the 
analogue gravity metric which describes the  effective geometry
of the expanding chiral fluid using 
the the formalism of relativistic acoustic geometry \cite{moncrief,bilic3,visser2}.
The equation of motion  (\ref{eq013})  may be written in the form
\begin{equation}
\frac{1}{\sqrt{-G}}\,
\partial_{\mu}
(\sqrt{-G}\,
G^{\mu\nu})
\partial_{\nu}
\mbox{\boldmath{$\pi$}}
+\frac{c_{\pi}^2}{a}V(\sigma,
\mbox{\boldmath{$\pi$}})
\mbox{\boldmath{$\pi$}}=0 ,
\label{eq028}
\end{equation}
with the analogue metric tensor,
its inverse, and its determinant given by
\begin{equation}
G_{\mu\nu} =\frac{a}{c_{\pi}}
[g_{\mu\nu}-(1-c_{\pi}^2)u_{\mu}u_{\nu}] ,
\label{eq022}
\end{equation}
\begin{equation}
G^{\mu\nu} =
\frac{c_{\pi}}{a}
\left[g^{\mu\nu}-(1-\frac{1}{c_{\pi}^2})u^{\mu}u^{\mu}
\right],
\label{eq029}
\end{equation}
\begin{equation}
G = \frac{a^4}{c_{\pi}^2}g .
\label{eq030}
\end{equation}
Hence, the pion field propagates in a (3+1)-dimensional
effective geometry described by the metric
$G_{\mu\nu}$.
It is convenient to work in comoving coordinates $(\tau,y,\vartheta,\varphi)$
with
 background metric $g_{\mu\nu}$ defined by (\ref{eq108}).
In these coordinates 
the analogue metric tensor (\ref{eq022}) is diagonal with components
\begin{equation}
G_{\mu\nu}=\frac{a}{c_\pi}{\rm diag} \left(c_\pi^2,  -\tau^2, -  
\tau^2\sinh^2\! y, - \tau^2\sinh^2\! y \sin^2\!\theta \right),
\label{eq008}
\end{equation}
where the parameters $a$ and $c_\pi$ are functions of the temperature $T$ which in turn
is a function of $\tau$  by (\ref{eq007}). In the following we assume
that these functions are positive.

In contrast to \cite{bilic3}, where it was assumed that both the background geometry and 
 the  flow were stationary, in an expanding fluid the flow 
is essentially time dependent. Hence, 
the acoustic geometry formalism  
 must be adapted to a non-stationary spacetime.
To this end we  introduce the concept of 
analogue marginally trapped surface or 
analogue apparent horizon.


To define the analogue apparent horizon 
we need to examine the behaviour of radial null geodesics of the analogue metric
(\ref{eq008}) in which $a$ and $c_\pi$ are functions of $\tau$. 
 In the following we assume spherical symmetry and denote by 
$l_+^\mu$ and $l_-^\mu$ the vectors tangent to outgoing and ingoing 
affinely parameterized 
radial null geodesics 
normal to a spherical two-dimensional surface. The tangent vectors are null with respect to the metric
(\ref{eq022}), i.e.,
\begin{equation}
G_{\mu\nu}l_+^\mu l_+^\nu =G_{\mu\nu}l_-^\mu l_-^\nu= 0. 
\label{eq143}
\end{equation}
Using the geodesic equation
\begin{equation}
l^\mu \nabla_\mu{l^\nu}=0,
\label{eq009}
\end{equation}
where the symbol $\nabla_\mu$  denotes a covariant derivative associated with
the metric (\ref{eq022}),
one easily finds  the tangent null vectors  corresponding to future directed  radial null geodesics,
\begin{equation}
l_\pm^\mu  = \frac{1}{a\tau}\left( 1, \pm\frac{c_\pi}{\tau},0,0\right),
\label{eq010}
\end{equation}
The null vectors
 $l_+^\mu$ and $l_-^\mu$ point towards increasing and decreasing $y$, respectively.
Hence, we adopt the usual convention and refer 
to $l_+^\mu$ and $l_-^\mu$ (and the corresponding null geodesics) as outgoing and ingoing.

The key element in the study of trapped surfaces is the 
expansion parameter $\varepsilon_\pm$ of  null geodesics.
A two-dimensional surface $S$ with spherical topology is called a {\em trapped surface}  if 
the families of ingoing and outgoing null geodesics normal to the surface are  both converging or both diverging.
More precisely,  the 
 expansion parameters
\begin{equation}
\varepsilon_\pm=\nabla_\mu l_\pm^\mu  
\label{eq244}
\end{equation}
 on a trapped surface $S$ should satisfy
$\varepsilon_+\varepsilon_- >0$.
A two-dimensional surface $H$ is said to be {\em future inner marginally trapped} 
if the future directed null expansions 
 on $H$ satisfy the conditions:  $\varepsilon_+|_H=0$, $l_-^\mu\partial_\mu\varepsilon_+|_H>0$ 
 and $\varepsilon_-|_H<0$.
 We shall refer to this surface  as
the {\em apparent horizon} since it is equivalent to the apparent horizon in cosmological context.

From (\ref{eq244}) with (\ref{eq010}) we find
\begin{equation}
\varepsilon_\pm  =\frac{2}{a\tau^2}
\left(\frac{\partial_\tau \left(\tau\sqrt{a/c_\pi}\right)}{\sqrt{a/c_\pi}}\pm  \frac{c_\pi}{v}  \right) .
\label{eq110}
\end{equation}
From  (\ref{eq110})  one finds the condition for the apparent horizon
 \begin{equation}
\frac{c_\pi}{v} \pm \frac{\partial_\tau \left(\tau\sqrt{a/c_\pi}\right)}{\sqrt{a/c_\pi}} =0.
\label{eq118}
\end{equation}
This equation defines a  hypersurface which we refer to as the {\em analogue  trapping horizon}.
The condition (\ref{eq118}) provides a functional relation between $y$ and $\tau$: 
\begin{equation}
\tanh y = f(\tau)\equiv \pm \sqrt{a c_\pi}/\partial_\tau \left(\tau\sqrt{a/c_\pi}\right),
\label{eq11}
\end{equation}
which by the coordinate transformation (\ref{eq147}) yields an implicit functional dependence of $r$ on $t$:
\begin{equation}
r/t=f(\sqrt{t^2-r^2}).
\label{eq2}
\end{equation}
Any solution to this equation, e.g.,  in terms of $r$ for an arbitrary fixed $t=t_*$,
 gives the location of 
the analogue apparent horizon $r_H$.


The critical behaviour of the null expansions $\varepsilon_+$ and $\varepsilon_-$
and of the derivative $l_-^\mu\partial_\mu\varepsilon_+$  near the trapping horizon is of particular interest.
Using (\ref{eq251})-(\ref{eq253}) in the neighbourhood of the horizon we find 
\begin{equation}
\varepsilon_+  \simeq\frac{2}{a\tau_{\rm c}^2}
\left(\frac{c_\pi}{v} -\frac{\eta\tau_{\rm c}}{\tau-\tau_{\rm c}}    \right),
\label{eq111}
\end{equation}
\begin{equation}
\varepsilon_- \simeq -\frac{4\eta}{a\tau_{\rm c}(\tau-\tau_{\rm c})}  ,
\label{eq211}
\end{equation}
\begin{equation}
l_-^\mu\partial_\mu\varepsilon_+ \simeq 
\frac{(d-2)\nu\eta + 2\eta (1+\eta)}{a^2\tau_{\rm c}^2(\tau-\tau_{\rm c})^2} ,
\label{eq215}
\end{equation}
where  
\begin{equation}
\eta=\beta-(d-2)\nu/4 .
\label{eq213}
\end{equation}
and $\tau_{\rm c} = c_0/T_{\rm c}$ is the
critical proper time corresponding to the critical temperature $T_{\rm c}$.
 Note that for an arbitrary fixed $t=t_*$ the outgoing null expansion $\varepsilon_+$
 vanishes at the radius $r=r_H$
at which $v=c_\pi (\tau-\tau_{\rm c})/(\eta\tau_{\rm c})$. 
The ingoing null expansion is obviously negative whereas the derivative of $\varepsilon_+$
given by (\ref{eq215}) is positive at $r_H$ and, 
according to the standard convention \cite{hayward},
 the region $\{r>r_H, t=t_* \}$ is future trapped with the apparent horizon located
 at  $r_H$ as its inner boundary.

At this point, it is worthwhile  examining the effective Hubble parameter
in the neighbourhood of $\tau_{\rm c}$.
The effective Hubble parameter for the  spacetime defined by the metric (\ref{eq008})
is given by
\begin{equation}
{\cal H} =\frac{\partial_\tau \left(\tau\sqrt{a/c_\pi}\right)}{a\tau}
\label{eq3}
\end{equation}
Using (\ref{eq252}) and (\ref{eq253}) with (\ref{eq007}) one finds the scaling behavior of ${\cal H}$ in the 
neighbourhood of $\tau_{\rm c}$:
\begin{equation}
{\cal H} \propto - (\tau-\tau_{\rm c})^{-1+\beta-3(d-2)\nu/4}.
\label{eq4}
\end{equation}
With the critical exponents $\nu=0.73$ and $\beta=0.38$
in $d=3$ dimensions 
 we find ${\cal H} \propto-(\tau-\tau_{\rm c})^{-1.17}$,
so the effective Hubble parameter is negative and diverges 
as $\tau$ approaches $\tau_{\rm c}$. Hence, our analogue spacetime describes
a shrinking universe with a singularity at the critical point.

\section{Surface gravity and analogue  Hawking effect}
\label{surface}

 On suggestion by Hajicek \cite{hajicek} it was recently argued
\cite{fodor,nielsen1,nielsen2,faraoni,pielahn,cai} that the Hawking effect
might be associated with the apparent horizon rather than the event horizon.
This observation is irrelevant for a stationary spacetime where the apparent and event horizons coincide.
In this case the apparent horizon is Killing and the surface gravity is uniquely defined
as a parameter that measures the inaffinity of the properly normalized Killing vector $\xi^\mu$.
The surface gravity $\kappa$ of the Killing horizon can be defined by
\begin{equation}
\xi^\nu\nabla_\nu\xi_\mu=\kappa \xi_\mu ,
\label{eq223}
\end{equation}
evaluated on the horizon. 
If the geometry were stationary, i.e.,
if  the components of  $G_{\mu\nu}$ in (\ref{eq022})
 were time independent,
 the apparent horizon would coincide with the chiral event horizon 
at the surface defined by $v=c_\pi$.
In that case
the surface gravity would read \cite{bilic3}
\begin{equation}
\kappa=\frac{1}{1-c_{\pi}^2}
\frac{\partial}{\partial r}
(v-c_{\pi}),
\label{eq043}
\end{equation}
where 
 the derivative  is to be taken
 at the horizon.

In the case of nonstationary spacetime, the apparent horizon is neither Killing nor null.
The definition of surface gravity in this case is not unique
\cite{nielsen2}
 and
several ideas have been put forward how to generalize the definition of surface gravity for the apparent horizon
\cite{hayward2,hayward,fodor,mukohyama-booth}.
In this paper we adopt the  prescription of 
\cite{hayward2} which, we believe, is
 most suitable for spherical symmetry.
This prescription involves the so-called Kodama vector $K^\mu$ \cite{kodama} which generalizes 
the concept of the time translation Killing vector to non-stationary spacetimes.
The Kodama vector we define as \cite{hayward,abreu}
\begin{equation}
K^\alpha= k \epsilon^{\alpha\beta}  n_\beta \;\; {\rm for}\; \; \alpha=0,1; \hspace{1cm}
K^i=0\;\; {\rm for} \;\; i=2,3,
\label{eq224}
\end{equation}
where $\epsilon^{\alpha\beta}$ is the covariant 
two-dimensional Levi-Civita tensor in the 
space normal to the surface of spherical symmetry
and  $n_\alpha$ is a vector normal to that  surface.
For the metric (\ref{eq008}) $n_\alpha$ is
given by 
\begin{equation}
n_\mu=\partial_\mu \left(\tau\sqrt{\frac{a}{c_\pi}} \sinh y\right).
\label{eq114}
\end{equation}
Our definition differs from the original one \cite{hayward,abreu}
 by   a normalization factor $k$ 
which we have introduced in order to meet the requirement that $K^\mu$
should coincide with the time translation Killing
vector $\xi^\mu$  for a stationary geometry.
Using the definition (\ref{eq224}) with metric (\ref{eq008})  we find
\begin{equation}
K^\tau= \frac{k}{\sqrt{a c_\pi}}\cosh y ; \hspace{1cm} K^y=- k \frac{\partial_\tau\left(\tau\sqrt{a/c_\pi}\right)}{\tau a} \sinh y ,
\label{eq225}
\end{equation}
with norm squared
\begin{equation}
|K|^2\equiv h_{\alpha\beta}K^\alpha K^\beta = {k^2}\sinh^2 y
\left[\coth^2 y - \left(\frac{\partial_\tau \left(\tau\sqrt{a/c_\pi}\right)}{\sqrt{ac_\pi}}\right)^2
\right] .
\label{eq226}
\end{equation}
The vector $K^\alpha$ is spacelike in the trapped region and vanishes on the trapping horizon.

In order to fix $k$, consider a general spherically symmetric analogue  geometry with
 metric ({\ref{eq022}).
In $(t,r)$ coordinates
\begin{equation}
n_\mu=\partial_\mu \left(r\sqrt{a/c_\pi} \right).
\label{eq214}
\end{equation}
Using 
 (\ref{eq224}) with metric  ({\ref{eq022}) we find
 \begin{equation}
K^t=k a^{-1}\partial_r \left(r\sqrt{a/c_\pi}\right); \hspace{1cm} K^r=-ka^{-1}r\partial_t \sqrt{ a/c_\pi} .
\label{eq325}
\end{equation}
In the case of stationary geometry
$K^r=0$ because
the quantities $a$ and $c_\pi$ depend only on $r$. Then, if we set
\begin{equation}
k=\sqrt{ac_\pi}\left(1+\sqrt{c_\pi/a}r\partial_r \sqrt{ a/c_\pi} \right)^{-1},
\label{eq326}
\end{equation}
  we  have 
  $K^\mu = \xi^\mu$ in the stationary case.
In $(\tau,y)$ coordinates
  \begin{equation}
k= \sqrt{a c_\pi} \left(\cosh^2y- 
\frac{\partial_\tau \left(\tau\sqrt{a/c_\pi}\right)}{\sqrt{a/c_\pi}} \sinh^2y \right).
\label{eq324}
\end{equation}

In analogy with (\ref{eq223}) the surface gravity $\kappa$ is defined by  \cite{hayward2,hayward3}
\begin{equation}
K^\alpha \nabla_{[\alpha} K_{\beta]}=\kappa K_\beta ,
\label{eq227}
\end{equation}
or equivalently  by
\begin{equation}
\kappa =\frac{1}{2} \frac{1}{\sqrt{-h}} \partial_\alpha ( \sqrt{-h}h^{\alpha\beta}k n_\beta),
\label{eq228}
\end{equation}
where  the quantities on the right-hand side should be evaluated on the trapping horizon.
Using 
({\ref{eq214}) and ({\ref{eq326})
it is easy to check that  this expression reduces to 
(\ref {eq043}) for a stationary spherical flow, i.e., a steady flow where
the velocity $v$ and the constants $a$, $b$, and  $c_\pi$ are time independent. 
By making use of the horizon condition
(\ref{eq118}) for our time dependent Bjorken spherical  flow
 we find from (\ref {eq228})
\begin{equation}
\kappa =\frac{\sqrt{1-v^2}}{(1+c_\pi v)^2}
\left[
\frac{c_\pi}{2\tau v}
 + \frac{c_\pi^2 -c_\pi^2 v-c_\pi v^2}{\tau} -\frac{c_\pi^2 v^2}{2\tau}
+\frac{\partial_\tau^2 \left(\tau\sqrt{a/c_\pi}\right)v}{\sqrt{a c_\pi}} 
\right] ,
\label{eq231}
\end{equation}
where $v=\tanh y$ and it is understood that the right-hand side should be
evaluated on the trapping horizon.
In other words, the function $\kappa=\kappa(\tau,y)$ in (\ref{eq231}) is a function of
two dependent  variables $y$ and $\tau$  subject to the constraint (\ref{eq118}).
Hence, $\kappa$ is effectively a function of only one variable, e.g.,
$\kappa=\kappa(\tau,y(\tau))$ 
through the explicit dependence on $\tau$ and implicit dependence via $y(\tau)$.

 The critical behaviour of $\kappa$
is determined by the behaviour of the quantity $\sqrt{a/c_\pi}$ in the neighbourhood of
the critical point. From (\ref{eq251})-(\ref{eq253})
we have
\begin{equation}
\sqrt{\frac{a}{c_\pi}}\propto  (\tau-\tau_{\rm c})^{-\eta}
\label{eq045}
\end{equation}
in the neighbourhood of 
the critical point,
where $\eta$ is defined in (\ref{eq213}).
Then, on the trapping horizon in the limit  $\tau \rightarrow \tau_{\rm c}$ 
the two dominant terms in square brackets in  (\ref{eq231}) scale as
\begin{equation}
\frac{c_\pi}{2\tau v} = 
-\frac{\partial_\tau \left(\tau\sqrt{a/c_\pi}\right)}{2\tau\sqrt{a /c_\pi}} 
\simeq
\frac{1}{2} \eta (\tau-\tau_{\rm c})^{-1},
\label{eq046}
\end{equation}
\begin{equation}
\frac{\partial_\tau^2 \left(\tau\sqrt{a/c_\pi}\right)v}{\sqrt{a c_\pi}}
\simeq (\eta+1)(\tau-\tau_{\rm c})^{-1}.
\label{eq047}
\end{equation}
Plugging these expressions in (\ref{eq231})
we find that the surface gravity in the neighbourhood of $\tau_{\rm c}$ behaves as 
\begin{equation}
\kappa \simeq (\eta+1/2)(\tau-\tau_{\rm c})^{-1} .
\label{eq048}
\end{equation}
From (\ref{eq213}),
with the critical exponents $\nu=0.73$ and $\beta=0.38$ for  the O(4) universality class
in $d=3$ dimensions \cite{baker}, we find $\eta=0.1975$.

The  temperature 
\begin{equation}
T_H=\frac{\kappa}{
 2\pi}
\label{eq044}
\end{equation}
associated with  the analogue apparent horizon is
the analogue Hawking temperature of thermal pions emitted at the apparent horizon
as measured by an observer near
 infinity. Since the background geometry is flat, this temperature  equals
 the locally measured Hawking temperature at the horizon.
From (\ref{eq048}) it follows that the analogue Hawking temperature may be arbitrary large
in the limit when the analogue horizon approaches the critical point.
However, Eq. (\ref{eq048}) holds for an ideal spherical expansion of the chiral fluid
and  in a realistic scenario one could only expect that the Hawking temperature 
may be  comparable with  or larger than the background fluid temperature.



\section{Discussion}
\label{conclusion}

It is tempting to speculate about possible signals for the analogue Hawking effect in
a  hadronic fluid.
In principle, one could  measure the temperature by fitting the 
pion spectrum to the thermal  Planck distribution.
However, one must be cautious before making any prediction for a realistic physical
system. First, one must invent a reliable signal to distinguish between the thermal pions produced above
 the critical temperature from those emitted as an analogue Hawking radiation from the apparent horizon
 below the critical temperature. Second,
a spherically symmetric expansion model considered here is not realistic
 for high energy heavy ion collisions.
A more realistic hydrodynamic model would involve 
a transverse expansion
superimposed on a longitudinal boost invariant expansion.
In this case the calculations become rather involved as the formalism for general nonspherical
spacetimes is not yet fully developed. This work is in progress.

\subsection*{Acknowledgments}
This work was supported by the Ministry of Science,
Education and Sport
of the Republic of Croatia under Contract No. 098-0982930-2864.


\end{document}